# Conductivity oscillations in 2D superlattice with non-harmonical dispersion law under non-quantizing electric and magnetic fields ( in Russian )

**G.M.Shmelev**

Volgograd State Pedagogical University, Volgograd, 400131, Russia

**Abstract**

*We calculate the current density in a semiconductor superlattice with parabolic miniband under crossed non-quantizing electric and magnetic fields. The Corbino disk geometry is considered. The current-voltage curve contains oscillations with period proportional to the magnetic field. The possibility is shown of the negative absolute conductivity. The Ampere-Gauss characteristics also contain overshoots under high enough electric fields. In all cases, the peaks smear with temperature rising.*

# Осцилляции проводимости двумерной сверхрешетки с негармоническим даконом дисперсии электронов в неквантующих Электрическом и магнитном полях

**Г.М. Шмелев**

Волгоградский Государственный Педагогический Университет,

400131, Россия

**Аннотация**

*Представлены результаты расчета тока в невырожденном квадратно модулированном двумерном электронном газе( 2МЭГ) с «параболической минизоной», находящемся в скрещенных неквантующих электрическом и магнитном (Е и Н) полях. Найдены компоненты тока вдоль главных осей ( OX и OY ) квадратной решетки при условии равенства нулю поперечного ( холловского) электрического поля ( геометрия Корбино ) . При $H \neq 0$ вольтамперная характеристика является мульти-N-образной с осцилляциями, период которых пропорционален H , при этом возможно появление областей с абсолютной отрицательной проводимостью.В достаточно сильном тянущем электрическом поле ток вдоль OY и гаусс-амперная тхарактеристика также содержат повторяющиеся пики, резкость которых уменьшается с ростом температуры*

## 1. Введение

В последние годы активно обсуждается надежда создания на основе полупроводниковых сверхрешеток ( СР ) терагерцового блоховского генератора с управляемой постоянным электрическим полем частотой[ 1,2 ]. Одним из основных препятствий на пути реализации такого генератора является низкочастотная элекорическая доменная неустойчивость. Подавить ее можно ,используя, в частности , СР с негармоническим законом дисперсии, в которых эффективная масса электрона положительна в большей части минизоны и растет к её краям .. В таких СР по сравнению с «традиционными» СР с синусоидальной минизоной статическая ОДП, являющаяся причиной вышеуказанной неустойчивости, сдвигается в сторону сильных полей, а высокочастотная – в сторону слабых. Более того, если ширина минизоны такой СР близка к энергии оптического фонона, то реализуется односторонний или двусторонний стриминг, и область статической ОДП дополнительно сдвигается в сторону больших электрических полей или совсем исчезает Таким образом, области динамической и статической ОДП в таких СР оказываются разнесенными по величине постоянного электрического поля, и возникает возможность генерации ( усиления ) терагерцового поляна участках ВАХ с *положительной* статической дифференциальной проводимостью. Добавим что еще в « пионерских» работах[3,4 ] обсуждался ( помимо синусоидального ) и закон дисперсии, составленный из сшитых прямой и перевернутой парабол . Частным случаем такой модели является и рассматриваемый в настоящей работе параболический закон дисперсии.  Возникает принципиальный вопрос: , реалистична ли модель СР с параболическим законом дисперсии? Положительный ответ  содержится в работе [5], где показано, что  данный закон дисперсии может быть реализован ,в частности, в СР со сложной элементарной ячейкой сверхрешеточного потенциала,  содержащей двойную несимметричную квантовую яму. В этой же работе, а также в [ 6 ],  предложены конкретные варианты реальных СР на основе $GaAsAl_xGa_{1-x}As$ , в которых может иметь место данный закон дисперсии. Короче говоря, здесь предполагается, что  вплоть до границы зоны Бриллюена закон дисперсии электрона квадратичный (параболический) или квазипараболический. Еще обратим внимание на работы [7,8] , где теоретически изучались некоторые свойства модели СР с параболической минизоной

## 2. Постановка задачи и результаты

Квадратичный закон дисперсии электрона в схеме расширенных зон можно записать в виде

$$\varepsilon(\vec{p}) = \varepsilon(p_x) + \varepsilon(p_y) ,$$

$$\varepsilon(p_{x,y}) = p_{x,y}^2 / 2 - 2\sum_{n=-\infty}^{\infty}(p_{x,y} - 2n - 1)[\theta(p_{x,y} - 2n - 1) - \theta(-2n - 1)] , \qquad (1)$$

где квазиимпульс ($\vec{p}$) и энергия ($\varepsilon$) представлены в единицах $\pi\hbar/d$ и $\Delta$, $\Delta/2$ - ширина минизоны, $d$ – период СР, $\theta(x)$ - ступенчатая функция Хевисайда.
Скорость электрона $\partial\varepsilon/\partial\vec{p}$ ( в единицах $\Delta d/\pi\hbar$ ) равна

$$v_{x,y} = p_{x,y} - 2\sum_{n=-\infty}^{\infty}\theta(p_{x,y} - 2n - 1) - \theta(-2n - 1), \qquad (2)$$

здесь второе слагаемое описывает скачки скорости при брэгговских отражениях на границах зон Бриллюена . В промежутках между этими отражениями электрон движется под действием постоянных неквантующих электрическом (**E** ) **и** магнитном ( вектор**H** направлен перпендикулярно **X**OY-плоскости 2DCP, ось ОХ направляем вдоль вектора **E** ).При $E_y = 0$ ( геометрия диска Корбино )
$p_x(t) = r\sin(\omega t); p_y(t) = r(\cos(\omega t) - 1); r = E/\omega$, нулевые начальные условия соответствуют нулевой температуре, $\omega$ - циклотронная частота.

Следуя методу Чамберса [ 8 ], подставляем эти выражения в ( 2 ) и усредняем скорость по времени с множителем $\exp(-t/\tau)$, $\tau$ - время релаксации импульса, В результате расчета , детали которого имеются в [7] ,получаем формулы для плотности тока ( в единицах $en\Delta d/\pi\hbar$, $n$ -плотность носителей):

$$j_x = \frac{E}{1+\omega^2} - 2ch^{-1}(\frac{\pi}{2\omega})\sum_{n=0}^{s}\theta(\frac{E}{\omega} - 2n - 1)sh[\arccos(\frac{\omega}{E}(2n+1))/\omega], \qquad (3)$$

$$j_y = -\frac{E\omega}{1+\omega^2} + 2sh^{-1}(\frac{\pi}{\omega})\sum_{n=0}^{s}\theta(\frac{2E}{\omega} - 2n - 1)sh[2\arccos(\sqrt{\frac{\omega(2n+1)}{2E}})/\omega] , \qquad (4)$$

при этом переобозначено : $\omega\tau \to \omega$, $eEd\tau/\pi\hbar \to E$ ; s –целая часть максимального значения первого слагаемого в аргументах $\theta$ -функции. В не слишком слабых магнитных полях падающая часть ВАХ $j_x(E)$ содержит обусловленные штарк-циклотронным резонансом осцилляции, период которых пропорционален Н . На рис. 1 представлена ВАХ для различных значений магнитного поля ($\omega$ =0.3,а; 0.6,б; 1,в). Последнее стимулирует рост тока в максимумах ВАХ (отрицательное магнетосопротивление ) и приводит к появлению областей с абсолютной отрицательной проводимостью. На рис.2 –плотность тока в холловском направлении $j_y(E)$: $\omega$ = 0.05,а; 1,б; 4,в. Гаусс-амперная характеристика также содержит всплески. Отметим , что во всех случаях рост температуры сглаживает резкость пиков. Для параметров $\Delta = 0.02 eV, d = 5.10^{-7} cm, \tau = 5.10^{-12} s$ величина $E = 1$ соответствует напряженности электрического поля $750 V/cm$, а $\omega = 1$ - $H = 10^4 Oe$.

## ЛИТЕРАТУРА

1. Ю.А.Романов . ФТТ,(2003), т.45, в.3,с.529 -533

**2**.Ю. А. Романов, Ю. Ю. Романова. ФТП 39, 1, 162 (2005).
**3.** L Esaki, R. Tsu, IBM J. Res. Dev. **14**, 2664 (1970)
4.P.A. Lebwohl, R. Tsu, J. Appl. Phys. **41**, 2664 (1970 )
5.Ю.А.Романов, Ю.Ю.Романова, И.В. Келейнов, Труды сканирующей зондовой микроскопии-2006. Нижний Новгород. 12 марта - 17 марта. 2006. c. 433-434
6. Ю.Ю. Романова¶, М.Л. Орлов, Ю.А. Романов.ФТП,   2012, том 46, вып. 11, с 1475-1483
7.G.M.Shmelev, E.M.Epshtein, M.B.Belonenko. arXiv:0905.3457v2 [cond-mat.mess-hall] ,May009
8     G.M. Shmelev, I.I. Maglevanny,
T.A. Gorshenina, E.M. Epshtein .    , J. Phys. A: Math. Theor. **41**, 501 (2008).

.

## Рисунки к статье Г.М.ШМЕЛЕВА

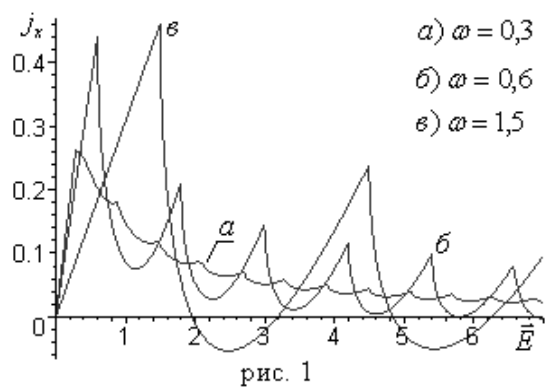

рис. 1

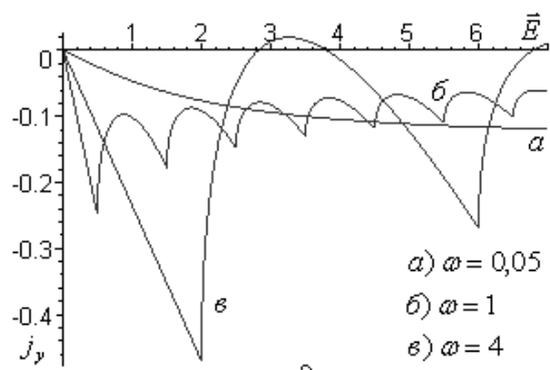

рис. 2

а) ω = 0,05
б) ω = 1
в) ω = 4